\documentclass[aps,pre, twocolumn, superscriptaddress]{revtex4}
\usepackage{graphicx}
\usepackage{dcolumn}
\usepackage{verbatim}

\newcommand{\be}{\begin{equation}}
\newcommand{\bea}{\begin{eqnarray}}
\newcommand{\ee}{\end{equation}}
\newcommand{\eea}{\end{eqnarray}}

\def\x{\mathbf{x}}
\def\y{\mathbf{y}}
\def\p{\mathbf{p}}
\def\e{\mathbf{e}}
\def\DF{\mathbf{DF}}
   
\begin{document}

\title{Parameter and State Estimation of Experimental Chaotic Systems Using Synchronization}

\author{Jack C. Quinn}
\email{jquinn@ucsd.edu}
\affiliation{Department of Physics, University of California, San Diego, La Jolla CA}

\author{Paul H. Bryant}
\email{pbryant@ucsd.edu}
\affiliation{Institute for Nonlinear Science, University of California, San Diego, La Jolla CA}

\author{Daniel R. Creveling}
\email{dcreveling3@mail.gatech.edu}
\affiliation{School of Physics, Georgia Institute of Technology, Atlanta, GA}

\author{Sallee R. Klein}
\affiliation{Department of Physics, University of California, San Diego, La Jolla CA}

\author{Henry D.I. Abarbanel}

\email{habarbanel@ucsd.edu}
\affiliation{Department of Physics, University of California, San Diego, La Jolla CA}
\affiliation{Institute for Nonlinear Science, University of California, San Diego, La Jolla CA}
\affiliation{Marine Physical Laboratory (Scripps Institution of Oceanography), University of California, San Diego, La Jolla CA}

\date{\today}
\pacs{05.45.Xt, 05.45.Tp}

\begin{abstract}
We examine the use of synchronization as a mechanism for extracting parameter and state information from experimental systems.  We focus on important aspects of this problem that have received little attention previously, and we explore them using experiments and simulations with the chaotic Colpitts oscillator as an example system.  We explore the impact of model imperfection on the ability to extract valid information from an experimental system.  We compare two optimization methods: an initial value method and a constrained method.  Each of these involve coupling the model equations to the experimental data in order to regularize the chaotic motions on the synchronization manifold.  We explore both time dependent and time independent coupling.  We also examine both optimized and fixed (or manually adjusted) coupling.  For the case of an optimized time dependent coupling function $u(t)$ we find a robust structure which includes sharp peaks and intervals where it is zero.  This structure shows a strong correlation with the location in phase space and appears to depend on noise, imperfections of the model, and the Lyapunov direction vectors.  For time independent coupling we find the counterintuitive result that often the optimal rms error in fitting the model to the data initially increases with coupling strength.  Comparison of this result with that obtained using simulated data may provide one measure of model imperfection.  The constrained method with time dependent coupling appears to have benefits in synchronizing long datasets with minimal impact, while the initial value method with time independent coupling tends to be substantially faster, more flexible and easier to use. We also describe a new method of coupling which is useful for sparse experimental data sets.  Our use of the Colpitts oscillator allows us to explore in detail the case of a system with one positive Lyapunov exponent.  The methods we explored are easily extended to driven systems such as neurons with time dependent injected current.  They are expected to be of value in nonchaotic systems as well.  Software is available on request.
\end{abstract}

\maketitle

\section{Introduction}

Physical models of nonlinear systems formulated as differential equations or as discrete time maps typically have unknown parameters representing our lack of detailed knowledge or our conjectures about how the dynamics operates in these systems. Further, forecasting the future of such systems requires the knowledge both of these parameters as well as the knowledge of the full state of the system at the time a forecast begins. Unfortunately, in interesting complex dynamics  as diverse as weather forecasting to prediction of networks of biological neurons, we cannot measure the full states of the system, and we must rely on an accurate parameter and state estimation procedure that allows reliable inference both of the parameters and of the full states of the dynamical systems from observations of some subset of the dynamical variables.

We make use of the phenomenon of chaos synchronization~\cite{pecora,Pecora2,kurths1,abar,kantz, Li, Uchida, Fotsin,kalnay}. This idea can be applied to the problem of parameter estimation by treating the measured data as the driver system, and implementing the model or response system numerically.  The receiver system is coupled to the driver system.  The fixed parameters and the initial conditions in the model/response system are adjusted to find the set of parameters which yields the smallest synchronization error of the two 
systems~\cite{ParlitzPRL, ParlitzPRE, Konnur, Huang, Maybhate, Sakaguchi, DRC}.   Sufficiently strong coupling can often mitigate the effects of a positive Lyapunov exponent in the response system.  It also tends to dramatically reduce the difficulty in finding a path through the parameter space from the initial guess to the optimal values.  This can often mean the difference between success and failure, because when the number of parameters is relatively large, it becomes a practical impossibility to fully explore that space.

In this paper we examine as a ``test system" a seemingly simple experimental system, the electronic Colpitts Oscillator. This system has the dual advantages of being easy to study, while still exhibiting difficulties in modeling that are commonplace in more complex systems such as networks of neurons.  It is often assumed that a perfect model for a system is available, but in reality no model is perfect and often the seemingly small imperfections of a model can have a substantial impact on our ability to extract accurate information about that system from the available data.  

In the control theory literature the task of determining the unobserved dynamical variables from observations of a subset of the whole collection of state variables goes under the name of finding observers for the system ~\cite{Nijmeijer, Nijmeijer2001, Nuijberts, Dedieu, Ciccarella93, Ciccarella95}. To the extent one may consider the unknown parameters of the system as state variables constant in time, the observer problem also includes the estimation of these parameters.

Chaotic oscillations of the system may mean that the observations we make and their equivalent dynamical variables in the models we develop for those nonlinear systems may not synchronize.  This will severely impede our ability to estimate either unknown states or parameters for the system of interest. This issue was first addressed by So, Ott, and Dayaswana~\cite{ottso94} in 1994 where they recognized that the states of the observed system, represented by vectors $\x(t)$, and the states of the model system, represented by vectors $\y(t)$, might not synchronize because of instabilities on the synchronization manifold $\x(t) \approx \y(t)$. They introduced an approach that looked at many of the ideas of observers from the point of view of this instability. They also recognized that the Kalman-like filters that are often lifted from their natural setting in linear problems to help achieve observability or synchronization in nonlinear problems are directed towards the reduction of noise in the tasks at hand rather than directed towards instabilities in the formulation of the problem itself.

We have introduced a variational formulation for state and parameter estimation in nonlinear problems that focuses on these issues as aspects of one problem. To illustrate the issues involved and the solutions we suggest, we have constructed an experiment on a simple, widely utilized, three variable nonlinear circuit, and we present both the formulation of our state and parameter estimation in that context as well as presenting the results of our analysis of the observed data for the circuit.

\begin{figure}
\begin{center}
\includegraphics[width=0.4\columnwidth]{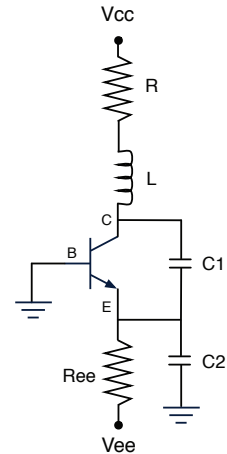}
\caption{\label{fig:circuit}Colpitts circuit}
\end{center}
\end{figure}

The circuit is shown in Fig.~\ref{fig:circuit}. It is a damped LRC resonator together with a bipolar junction transistor providing a nonlinear dynamical element. There are three independent dynamical variables for its description, and we choose them to be $V_{CE}(t)$, the voltage at the collector relative to the emitter, $V_{E}(t)$ the voltage at the emitter relative to ground, and $I_L(t)$ the current through the inductor.   

Our challenge in this paper is to use measurements of $V_E(t_n)$ taken every $t_n = n\Delta t ;\;\; \Delta t = 10\mu \,$s for $n = 0, 1, ..., N$ to estimate the parameters in the model of the circuit and to estimate the unobserved state variables $V_{CE}(t_n)$ and $I_L(t_n)$ over the same selection of observation times. Once we have estimated $V_{CE}(t_N), I_L(t_N)$ and the fixed parameters, we may use the model for prediction for $ t > t_N$.

It is the transistor in particular that makes this an interesting system for study. A simple transistor model is well known, but what is perhaps not very well known is that real transistors will follow this model only approximately.  As we will show, an imperfect model can have a significant impact on our analysis, especially on efforts to extract accurate parameter values.  We attempt to find methods to test the accuracy of our results and potential improvements to the model.  In carefully examining the effect of coupling strength between model and data we find interesting and {\bf anomalous} behavior which may provide a useful test for the quality of different models for the system.

We examine two methods for estimating parameters that build on the idea of stabilizing the synchronization manifold between the observations and the model dynamics ~\cite{abar2}. 

\begin{itemize}
  \item One method (initial value) determines the parameters and the state variables at the initial time $V_E(0), V_{CE}(0), I_L(0)$ by solving the model equations for a selection of parameters and initial conditions to produce a model counterpart of the observed variable's time series. A metric for the distance between the model output and the observations is then minimized using a standard procedure in initial condition, parameter space. The advantages of this method include speed and flexibility of use.  Software implementing this method is available from one of the authors (PB)~\cite{DataSync}.
  \item The other method (constrained optimization) works in the larger space of model variables at each observation time $t_n$, a coupling/control parameter at each of these times, and the time independent parameters. This method offers more efficient and precise control of synchronization, though it suffers from a number of practical problems, e.g. setup, speed, memory and numerical problems.  Software for implementing this method is also available~\cite{snctrl}.
 \end{itemize}
 
\subsection{Circuit Equations}
We work with a nonlinear oscillator of Colpitts variety. This has a long history in the study and utilization of nonlinear oscillators for a variety of technological applications. While it is likely that many engineers working with Colpitts and other nonlinear oscillators encountered chaotic oscillations, Kennedy~\cite{Kennedy} appears to be the first to have recognized this was not noise or an unwanted defect in circuit behavior. We utilize a chaotic Colpitts oscillator for illustrating our suggestions on estimating parameters and unobserved states in a chaotic system. It is important to move beyond the numerical study of these oscillators and demonstrate in an experimental setting how the ideas work, because theoretical models at best only approximately describe real experimental systems.   

We built the circuit shown in Fig.~\ref{fig:circuit} using the the following components: $C_1 = 7.44$ $\mu$F, $C_2 = 7.23$ $\mu$F, $L = 11.74$ mH, $R_{EE} = 392 ~\Omega$, and the power supply $V_{CC}= 5.03$ V, $V_{EE} = -5.10$ V.  The capacitors were non-electrolytic with a maximum voltage rating of at least 30V.  The inductor was air core to avoid any nonlinearity due to core saturation or hysteresis.  We used a 100 $\Omega$ potentiometer in the circuit to adjust the parameter $R$.  Note that $R$ in all the equations includes the resistance of the potentiometer plus the resistance of the inductor, which was not negligible in our case.  We used a 2N2222 BJT small signal transistor.  The fundamental frequency of this oscillator is approximately $f_0 = 1/(2 \pi \sqrt{L C_{eq}})$, where $1/C_{eq} = 1/C_1 + 1/C_2$.  For the circuit elements we used, $f_0 \approx 770$ Hz. 

The dynamics is described by three coupled first order differential equations, obtained directly from the circuit using Kirchoff's laws:
\be
C_1 \frac{ d V_{CE}(t) }{dt} = I_L(t) - I_C(V_E(t)),
\ee
\begin{equation}\label{dVEdt}
C_2 \frac{ d V_E(t)}{dt} = I_L(t) - \frac{V_E(t) - V_{ee}}{R_{ee}}   + I_B(V_E(t)),
\end{equation}
\begin{equation}\label{dILdt}
L \frac{d I_L(t)}{ dt} =  V_{cc} -V_E(t) - V_{CE}(t) - R I_L (t),
\end{equation}
where $I_C(V_E)$ and $I_B(V_E)$ are the currents into the collector and base of the transistor.  The state of the circuit is completely described by the three dynamical variables: $V_{CE}(t)$ the potential at the collector relative to the emitter, $V_{E}(t)$ the potential at emitter relative to ground, and $I_L(t)$ the current through the inductor.  A key ingredient needed to complete the description is the set of equations that specify the transistor currents.  Here we used a simplified version of the Ebers-Moll equations~\cite{emoll, Maggio},
\begin{equation}\label{IC}
I_C(V_E) = I_s \exp \left(\frac{-V_E}{V_T}\right),
\end{equation}
\begin{equation}\label{IB}
I_B(V_E) = \frac{I_C(V_E)}{\beta_F},
\end{equation}
where $\beta_F$ is the forward current gain and $I_s$ is the reverse saturation current, both properties of the particular transistor. $V_T = kT /e $ is the thermal voltage.  These equations are nonlinear and are what leads to the interesting behavior of the circuit.  Inaccuracy in the transistor model has important consequences for our analysis. Later in this paper we discuss some ways to improve it.

Although in this test circuit we can take experimental measurements of all three of the variables, it will often be the case in experiments on other systems that only some of the variables can be measured; this is certainly true, for example, of weather forecasting and nervous systems. So in our analysis we will assume that only $V_E(t)$ is available for analysis, and that the circuit parameters are unknown.   

\subsection{Circuit Dynamics}
The values of the circuit parameters determine the behavior of the circuit.  Although all the parameters matter, here we focus on changing only the resistance $R$ by adjusting the potentiometer, and keeping all the other parameters fixed. 

The simplest possible behavior of the circuit occurs when the three dynamical variables are constant in time; namely, a fixed point.   To find this point, set the time derivatives in the equations of motion to zero. The emitter current, $I_E = I_C+I_B$, can be found from Eqs.~(\ref{IC}) and (\ref{IB}), and set equal to the current through the resistor $R_{EE}$,
\begin{equation}
I_E = I_s(1+\frac{1}{\beta_F})\exp(\frac{-V_E^{(0)}}{V_T})=\frac{V_E^{(0)}-V_{ee}}{R_{ee}}.
\end{equation}
The fixed point emitter voltage $V_E^{(0)}$ can then be found numerically for specific parameters.   The other two fixed point values  $V_{CE}^{(0)}$ and $I_L^{(0)}$ can then be calculated, using $I_L^{(0)} = I_s\exp(-V_E^{(0)}/V_T)$, and $V_{CE}^{(0)} = V_{cc} -RI_L^{(0)} -V_E^{(0)}$.  The fixed point is a useful thing to calculate, because it can be used as a first step for comparing the model predictions to  the voltages measured in the actual circuit.  This fixed point is always a solution to the equations of motion, but it stable only when $R$ is large enough. 

\begin{figure}
\begin{center}
\includegraphics[width=1.0\columnwidth]{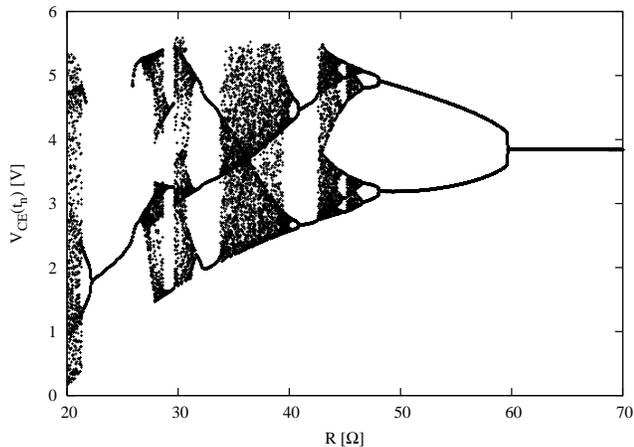}
\caption{\label{bifurcations} Bifurcation diagram calculated by numerically integrating the model equations over a range of values for $R$.  The value of $V_{CE}(t_n)$ is plotted for all $t_n$ where $V_E(t_n)$ = -0.6 V with $V_E$ increasing.  There is a period doubling bifurcation at $R \approx 59 \Omega$ and again at $R \approx 48 \Omega$.}
\end{center}
\end{figure}

As R is decreased the fixed point becomes unstable and the circuit oscillates at  approximately the fundamental frequency $f_0$.  As R is decreased further there is a series of period doubling bifurcations leading to chaos.  This is shown in the bifurcation diagram Fig.~\ref{bifurcations}.   To construct this diagram, we integrated the circuit equations using a range of values for $R$, and plotted the value of $V_{CE}$ only at the times $t_n$ where both $V_E(t_n)= V_{th}$ and $\frac{dV_E(t_n)}{dt}>0$.  The threshold value of $V_{th}=-0.6$ V was chosen because that is about where the transistor switches from ``on" to ``off".

\begin{figure}
\begin{center}
\includegraphics[width=0.85\columnwidth]{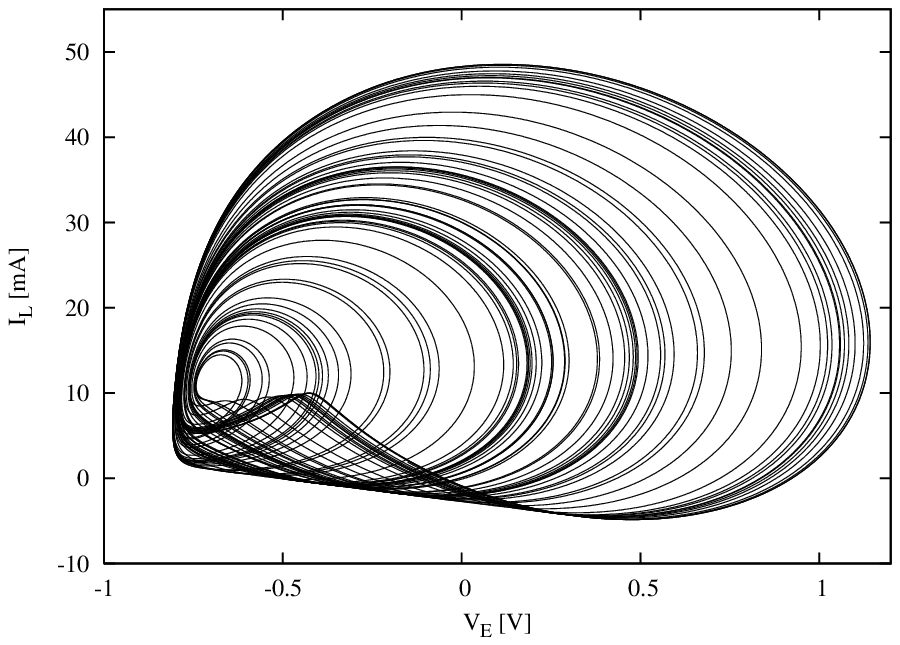}
\includegraphics[width=0.85\columnwidth]{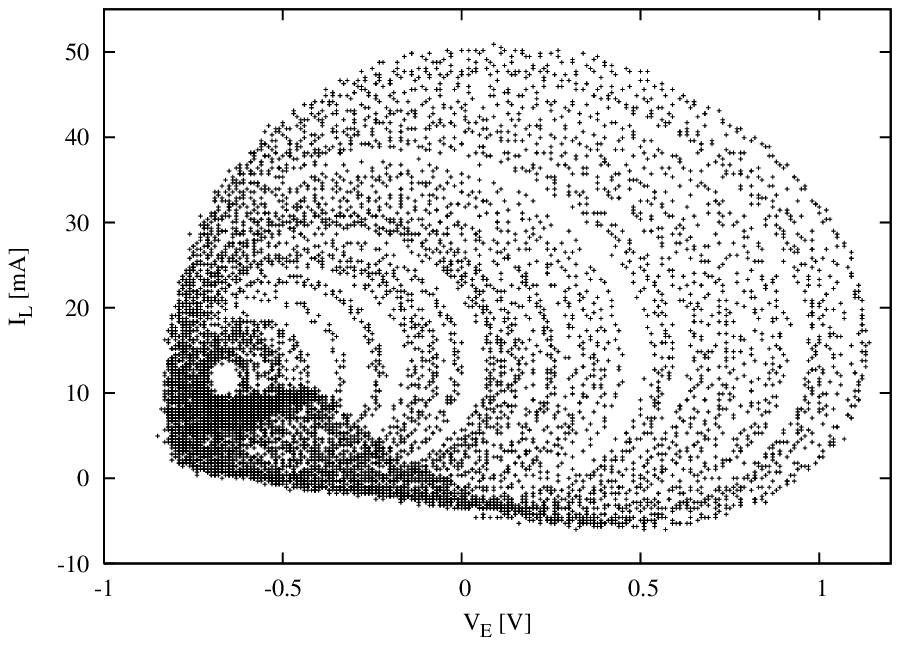}
\caption{Two dimensional view of the attractor: {\bf Top Panel} $V_E(t)$ versus $I_L(t)$ calculated by integrating the model equations;
{\bf Bottom Panel} $V_E(t)$ versus $I_L(t)$ as observed in our experiments.}
\label{twodmodelexpt}
\end{center}
\end{figure}

We recorded all three dynamical variables $V_{CE}(t), V_{E}(t), $ and $I_L(t)$  when the circuit was operating in a chaotic regime.  The orbit traces out a strange attractor in this three dimensional space.  A two dimensional projection of the attractor is shown in Fig.~\ref{twodmodelexpt} with data both from our experiments and from our model.   

\section{Parameter and State Estimation Using Synchronization}
\subsection{Background}
Although we were able to measure all the dynamical variables for our Colpitts oscillator, we use this as an example system to evaluate methods for estimating unknown parameters and unobserved state variables. We proceed as though we know many of the `external' parameters driving the circuit, such as the constant voltages $V_{ee}$ and $V_{cc}$ while treating the parameters within the circuit, especially those of the model used for the nonlinear operation of the transistor, as unknown. 

We proceed, then, with the assumption that we observe only $V_E(t_n)$ where $t_n = n \Delta t,$ with the data time-step $\Delta t = 10 \mu \,$s. We wish to use this information to evaluate the remaining parameters and the value of the state variables $V_{CE}(t_n)$ $V_E(t_n)$ and $I_L(t_n)$ for $n = 0$ through $N$.  By assumption we have data for $V_E(t_n)$, but this will be slightly different than the output from the model due to noise and imperfections in the model.  Except where otherwise specified, our results are for the case $N = 1000$ or $t_N = 10$ ms.  For all of the analyses in this paper, the model uses both a half-step and a full-step for improved accuracy.  Often the model half-step $h$ will be chosen to be equal to the data time-step. i.e. $h = \Delta t = 10 \mu \,$s.  In this case the full-step will be twice this value and the number of full-steps used in the analysis will be $N/2$.  As is discussed later in this paper it is sometimes advantageous to choose a smaller value for $h$, typically by selecting a value for which some multiple of $h$ is equal to $\Delta t$.

It is very important to choose an appropriate quantity of data to analyze.  With too little data there may be insufficient information to determine the unknown parameters, while with too much data, not only is the processing slowed down, but (as we will show) the problem of maintaining synchronization becomes more severe.  Once the fixed parameters are estimated along with $V_{CE}(t_N)$ and $I_L(t_N)$, we may use them and the observed $V_E(t_N)$ as initial conditions for the three dynamical circuit equations to predict $V_E(t > t_N)$. We would also predict $V_{CE}(t>t_N)$ and $I_L(t>t_N)$ at the same time, but in the scenario we are envisioning, those are not observable, so we would have no way to verify the predictions. The advantage of our experimental setup and our measurements of all state variables is that in the case we are discussing here, we do, in fact, know $V_{CE}(t_n)$ and $I_L(t_n)$ and can further examine the validity of our estimates. In the interesting physical settings we have in mind, this will not be the case, and we wish to have some sense of the reliability of the methods in that instance.

We make use of the phenomenon of synchronization of chaotic systems.  The key idea is that two coupled systems will synchronize the most readily when they are identical.  In our case, one system is the ``driver," which here is the actual circuit, and the other system is the ``response" or model system which is implemented numerically using the model equations with an additional term to couple to the driver system.  The response system has adjustable parameters, and our goal is to find the values for the parameters that best match the parameters of the driver system.  Here we introduce a simple form of coupling by taking the difference between a measured data variable and the corresponding model variable at each time step, multiplying this by a coupling strength $u$, and adding this expression to the right hand side of the corresponding differential equation.  This can be done for as many variables for which we have data, but since we are using Colpitts as a ``test system" we choose to only apply it to $V_E$.  This results in the following sets of equations, in which the primed variables represent our model of the system (response) and the unprimed variable $V_E$ represents the experimental data (driver):
\bea
\frac{ d V'_{CE} }{dt} &=& \frac{I'_L - I'_C(V'_E)}{C'_1}, \nonumber \\*
\frac{ d V'_E}{dt} &=& \frac{I'_L + I'_B(V'_E)}{C'_2} - \frac{V'_E - V'_{ee}}{C'_2 R'_{ee}} + u(V_E - V'_E),  \nonumber \\*
\frac{d I'_L}{ dt} &=&  \frac{V'_{cc} -V'_E - V'_{CE} - R' I'_L }{L'}.
\label{coupled}
\eea
Another possible coupling scheme (that we have not used) would be to replace $V'_E(t)$ with $\alpha V_E(t) + (1-\alpha)V'_E(t)$ on the right hand side of all three circuit equations, where $0 <\alpha <1 $. 

When model and data are different at some time-step, i.e. when $V_E(t) \ne V'_E(t)$, the coupling term will try to move $V'_E$ in the direction that closes this gap.  Note also that the coupling strength $u$ can be allowed to change with time-step or it can be independent of time-step.  It can also be put under the control of the optimization program, or it can be fixed, or it can be manually changed through a set of values.  We explore all of these options.  

Note that numerical instability is likely if $u$ is allowed to be too large.  The limiting value is approximately $1/2h$, i.e. one should require
\bea
u < 1/2h.
\label{unstable}
\eea
To increase $u$ beyond this point one can make a corresponding reduction in the model half-step, possibly requiring the experimental data to be interpolated.  Another common reason for reducing the timestep would be when it is necessary to improve the accuracy of the numerical integration of the equations when no coupling is involved.

An alternate scheme that avoids the need for interpolation of the data and that we have used quite successfully, is to remove the synchronization term from the differential equations, and instead make small synchronizing corrections directly to the model variable.  These corrections are applied between integration steps when there is data available at that time point, discontinuously moving the model variable some fraction of the way towards the corresponding experimental data value.  The small steps in the model variable will tend to vanish on approach to perfect synchronization.  This method of coupling can be shown to be essentially equivalent to the other method if that fraction is $1 - \exp(-u \Delta t)$.  Using subscripts $-$ and $+$ to represent before and after the discontinuous step, we have for our current problem:
\bea
V'_E(t_{n+}) = V'_E(t_{n-}) + (1 - \exp(-u \Delta t))(V_E(t_n) - V'_E(t_{n-}))
\label{syncstep}
\eea
\textbf{Note that this removes the stability problem for large $u$, i.e. Eq.~(\ref{unstable}) no longer applies.}  The case $u$ goes to infinity corresponds to a jump in the model variable all the way to the data value for that time-step.  One way to obtain Eq.~(\ref{syncstep}) is to leave the coupling term in the differential equation but replace $u$ with $u \Delta t \sum \delta (t - t_n)$, where the sum is over all data points (indexed by $n$), $\Delta t$ is the data time-step and $\delta()$ is the Dirac delta function.  This can be thought of as a special kind of time dependent coupling.  The delta function can be integrated out to give the fractional displacement rule above.  It is often the case that the desired timestep for the model is smaller than what is available or possible to achieve for the experimental data.  In cases like this where the data is \textbf{sparse}, the method just described allows the time-step of the data to be a multiple of the time-step of the model, conveniently solving this problem without needing to interpolate the experimental data and avoiding the additional inaccuracy that would be introduced by that interpolation.

We require a metric in which to compare our observed $V_E(t_n) = V_E(n) = x_1(n)$ with the estimates we generate from solving the model equations for $V'_E(t_n)=V'_E(n) = y_1(n)$. These estimates depend on the parameters, which we call $\p$ now, and the initial conditions $\y(0)=\{V'_E(0) = y_1(0), V'_{CE}(0) = y_2(0), I'_L(0) = y_3(0)\}$. We adopt the usual least squares or minimum variance metric as the distance between the experimental observations and the value of our model output
\begin{equation}\label{CostFn1}
C(\p,\y(0)) =\frac{1}{N+1}\sum_{n=0}^{N} (x_1(t_n) - y_1(t_n, \p,\y(0)))^2.
\end{equation}
We call this the cost function or objective function, and it is a measure of the quality of synchronization.  

\begin{figure}
\begin{center}
\includegraphics[width=1.0\columnwidth]{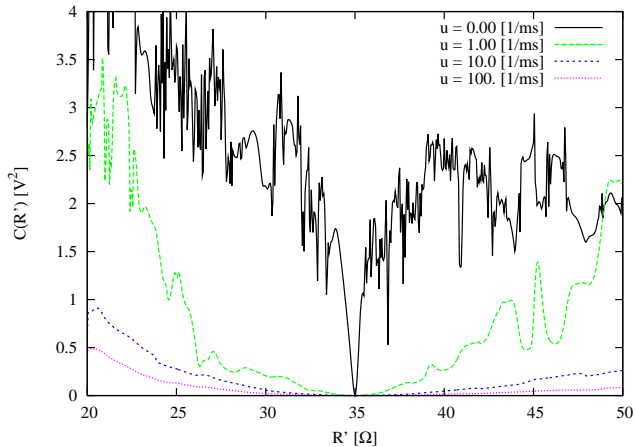} 
\caption{ Cost (Eq.~(\ref{costr'})) as a function of $R'$ for four different coupling strengths (decreasing from bottom to top).  The calculation was done over 10 ms, with the two systems starting at the same point in phase space, and all parameters identical except $R \neq R'$. Increasing $u$ smoothes out the surface, but it also makes the minimum less well defined as $u$ becomes large. }
\label{costofR}
\end{center}
\end{figure}

The goal is to find the parameters and initial conditions which minimize the cost function.  One problem is that the path through parameter space from the initial guess to the best result may contain local minima which will prematurely end the minimization process resulting in an incorrect result.  But even when this is not the case, the ``downhill" path through a high dimensional parameter space can sometimes be extremely convoluted and lengthy.  We can see an indication of this in Fig.~\ref{costofR} by looking at the variation of the cost function, $C(R')$, 
\begin{equation}
\label{costr'}
C(R') =\frac{1}{N+1}\sum_{n=0}^{N} (x_1(t_n) - y_1(t_n, R'))^2.
\end{equation}
when only the resistance $R'$ in the response circuit Eq.~(\ref{coupled}) is varied with all other parameters held fixed. The curve labeled $u = 0$ corresponds to integrating the model equations and evaluating the cost function Eq.~(\ref{CostFn1}). The other elements of Fig.~\ref{costofR} will be addressed in a moment.

This complicated behavior of $C(R')$ impedes one's ability to seek the global minimum where we expect that $R'$ will equal $R$ of the driver circuit--the source of our data.  As one might expect, as more parameters are added, the graph becomes a very complex hyper-surface that we must explore. This complexity is associated with the instability of the synchronization manifold $\x(t)= [ V_E(t), V_{CE}(t),I_L(t)] = \y(t) = [V'_E(t), V'_{CE}(t), I'_L(t)]$ in the six dimensional state 
space $[\x(t),\y(t)] = [V_E(t), V_{CE}(t),I_L(t), V'_E(t), V'_{CE}(t), I'_L(t)]$ of the coupled oscillators. This instability leads to extreme sensitivity of the orbit $\y(t)$ on its parameters, including here $R'$, and, as it happens on its initial conditions $\y(0)$ though the latter is not shown here~\cite{evensen}. This complexity of the surface of a cost function in parameter space has been known for some years~\cite{bruck,Kurths}.

It should be noted that when additional parameters are included, many (perhaps most) of the local minima shown in Fig.~\ref{costofR} will no longer be minima.  That is to say, in many cases it may be possible to escape from one of these apparent minima by simply moving in some orthogonal direction.  It also presents a somewhat oversimplified picture, in that the real experimental data will often include some significant noise, and also the model we are using to represent the experimental system may have significant imperfections.  As a result no choice of parameters will produce perfect synchronization and the optimal cost will be nonzero.

With these considerations in mind, now we examine the other curves in Fig.~\ref{costofR}. They correspond to increasing values of $u$, and we see two effects as $u$ is made larger: (1) the complex, rough nature of $C(R')$ as a function of $R'$ becomes smooth, and (2) the magnitude of $C(R')$ decreases. The origin of these two effects comes from the synchronization of the data $x_1(n) \to y_1(n,R')$. As $u$ becomes very large, it drives $x_1(n) \to y_1(n,R')$ approximately as $\frac{1}{u}$ leading to $C(R')$ decreasing as $\frac{1}{u^2}$.  Interestingly, for small or moderate size $u$, there is structure in the dependence of the cost function on $u$ that can reveal information about the quality of the model, the presence of noise, and other attributes useful in interpreting the outcome of the numerical optimization.  

The smoothing out of the surfaces in parameter and initial condition space is attributable to the reduction of any positive \textbf{conditional} Lyapunov exponents (CLEs)~\cite{pecora,abar,kurths1,kantz} by the added term $u(V_E(t) - V'_E(t))$.  Note that the term ``conditional" is used because the exponents depend on the experimental variable $V_E$ which is treated as an external drive signal whose dynamics are not explicity considered. The CLEs are evaluated by iterating the Jacobian $\DF(\y(t))$ along an orbits of the dynamical equations, and this change in the dynamics takes the original Jacobian and modifies it by subtracting $u$ from the (2,2) component in our example. This allows us to reduce the one positive CLE to a negative value, synchronize the data and the model output, and make the contributions to the cost function coherent.

\begin{figure}
\begin{center}
\includegraphics[width=1.0\columnwidth]{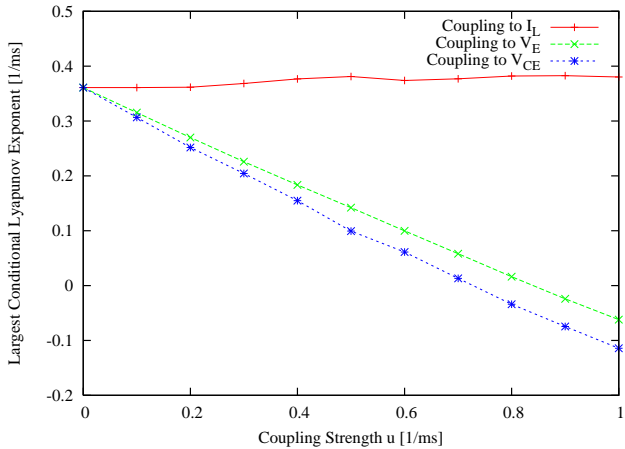} 
\caption{Largest conditional Lyapunov exponent (CLE) as a function of coupling strength. With coupling to $V_E$ or $V_{CE}$ the largest CLE is reduced from positive to negative, but for coupling to $I_L$ the largest CLE is actually increased.}
\label{CoupledLyap}
\end{center}
\end{figure}

From the point of view of the nonlinear dynamics expressed in Eq.~(\ref{coupled}), we are, by coupling the equations to the data, reducing the largest conditional Lyapunov exponent from positive to negative.  This idea is illustrated in Fig.~\ref{CoupledLyap}, which shows how the largest CLE depends on the coupling strength, and depends on which variable is being coupled. Coupling to $I_L(t)$ will not cause synchronization, at least for $u<1$.      

It is desirable to have $u \to 0$ at the end of any estimation procedure as the term $u(V_E(t) - V'_E(t))$ in our equations is not a result of Kirchoff's laws or any physics, but is introduced to regulate instabilities on the synchronization manifold. To use the equations for prediction, it should be gone from the dynamics.

Although increasing $u$ appears to make our parameter finding problem much easier it can also have an undesirable effect.  This comes from the possibility that in real experimental systems, where the model is imperfect, the coupling term in the equations of motion will not completely vanish for any choice of model parameters.  This in turn can cause the parameter values that minimize the cost function to be dependent on the magnitude of the coupling.  To deal with this, we can initially use a very strong coupling to arrive at a first approximation to the desired parameters, and then the coupling can be reduced gradually to smaller and smaller values, either to zero or to a value where its impact is small.  To accomplish this it is sometimes efficient to allow the optimization process to control the value of $u$.  In this case it may sometimes be necessary (though not always) to add a ``coupling cost" to the cost function Eq.~(\ref{costr'}) so that the cost can decrease as the coupling is decreased.
\begin{equation}
\label{costru}
C(\p,u) =\frac{1}{N+1}\sum_{n=0}^{N} \biggl \{(x_1(t_n) - y_1(t_n, \p)^2\biggr \} + \eta^2u^2,
\end{equation}
where $\eta$ is a number assigning a weight to the synchronization term relative to the coupling term, and $\p$ is a vector comprising all of the parameters and initial conditions. In some cases we allow the coupling to depend on time-step, based on the idea that the strength of the coupling required to make synchronization occur may vary across the system attractor.  One might expect that in certain locations the data variable being measured may sometimes be in good alignment with direction of maximal stretching of the attractor.  This alignment can be examined explicitly through the Lyapunov direction vectors (in Ref.~\cite{Brown} see discussion starting on page 2796).  This does of course complicate the optimization problem; if there are 500 full time-steps for the model, and we wish to optimize this $u(t)$ then we have just added an additional 500 unknown parameters to the problem.  We give results for both cases.

\subsection{Effect of Positive Lyapunov Exponents}

The Lyapunov exponents, particularly the most positive one, can be very useful in determining how to analyze the experimental data.  These measure the average rates of expansion or contraction of a chaotic attractor along different directions of the phase space.  There are methods for calculating these from experimental data~\cite{Bryant,Brown}, and other, generally easier, methods that can be used when a good model is available and the parameters are known (in Ref.~\cite{Eckmann1985} see discussion of QR decomposition method beginning on page 650 -- software implementing this method is available~\cite{LyapOde}).  When only the most positive exponent is needed, the problem is much easier -- nearby initial conditions (on the attractor) will separate on average as $\e^{\lambda t}$ , where $\lambda$ is the largest Lyapunov exponent.  For our system and parameter choices, we determined the most positive exponent to be about 0.35(1/ms).  For our choice of time interval $t_N=10$ ms for most of our results, one can expect that a small error in the initial condition at $t=0$ will grow by a factor $\e^{3.5}=33$ by the end of the interval.  As a result of this moderate growth over our interval, it is relatively easy to reduce the coupling to zero without degradation to the fit to the experimental data.  Going much beyond 10 ms, however will change this picture.  In order to reduce the coupling all the way to zero we need to require that
\begin{equation}
\label{IntervalLimit}
\lambda t_N < \Gamma,
\end{equation}
where $\Gamma$ is probably not a precise constant, but is rather a number that is roughly about 10 and which may depend somewhat on the details of the system and how hard we are willing to work on the analysis.
An increase to 100 ms definitely puts us over this limit for the present problem, drastically increasing the growth factor to $\e^{35}=1.6 \times 10^{15}$.  In this case it is quite impossible to reduce the coupling all the way to zero without major loss of synchronization.  We must choose to either reduce the time interval, or accept the problems that may possibly be caused by maintaining a nonzero coupling parameter.  In the latter case, having a time dependent $u(t)$ presents a clear benefit in being able to apply the coupling more efficiently at points where it is most beneficial.

\subsection{Optimization Methods}

We used two different optimization methods, an initial value method and a constrained method, which we describe below.  In both cases the primary goal was to find the best values for the parameters and all of the state variables based on the available experimental data which in this case is a time series for $V_E(t)$.  Once this has been achieved we can also make predictions by integrating the model into the future, beyond the last data point.  It turns out that for both of these methods, numerical accuracy of the model is improved by using a half time-step, at which equations are evaluated, while the full accuracy is achieved only at the full time-step points.  It was convenient for our data to use a half-step for the model that was equal to the time-step for the data.  So we use $t_n$ to represent the time points for the data as well as these half-step time points for the model.  The even values of $n$ correspond to the full-step time points.  However, for other experiments it will often be the case that this will \textbf{not} be a good choice.  In general, the model half-step should be chosen to be as large as possible without adversely effecting numerical accuracy.  But, often there are limitations on the sampling rate for experimental data that require that the data time-step be larger than this value, although usually one can arrange for the data time-step to be some multiple of the model half-step.

The first method we examined is an ``initial value method" similar to the one discussed in Voss et al.~\cite{Kurths}, but uses a time independent coupling term to synchronize the experimental data set to the model. This method treats the initial conditions $\y(0)$ as additional parameters to be determined.  The cost function can be computed at any point in this extended parameter space by numerically integrating the equations of motion with those values of the parameters, and comparing the result to the data using the cost function Eq.~(\ref{CostFn1}).  We use standard minimization methods to search the parameter space for the minimum of the cost function.  Integration of the equations is typically done using the standard fourth order Runge-Kutta algorithm.  Note that this algorithm, as mentioned previously, involves function evaluations at the midpoint of each full time-step.  Software (DataSync) implementing this initial value method is available from one of the authors (PB) on request~\cite{DataSync}.  

Two of the minimization methods that we tried with the initial value method were the Broyden-Fletcher-Goldfarb-Shanno (BFGS) ``quasi-Newton method" \cite{nr}, and the Brent-Zangwill-Powell-Smith (BZPS) ``direction set method" \cite{Brent}.  We first set upper and lower bounds and starting values for each of the parameters to be used by the minimization method.  We used physically reasonable values for bounds and starting values.  Most of the results in this paper were obtained using the BZPS method.  BZPS is convenient because it does not require evaluation of derivatives of the cost function.  The derivatives required by BFGS can however be determined numerically.  It has been our experience that BZPS is the most reliable method, even though it may sometimes be slightly slower than the BFGS method.  The BZPS method works by performing line searches to find the minima of the cost function along specific directions in parameter space.  Each direction is searched along in a specific sequence, always staring off at the location in parameter space where the last search ended.  Initially the search directions are just the cardinal directions, but as the process goes on the directions are updated based on the results of previous iterations.    

In this initial value method, we often set the coupling strength $u$ to be time independent and fixed at a user specified value.  We use the method in several steps. Initially we use a large $u$ to get from the initial guess of parameters to the correct region of the parameter space.  Then we repeat with smaller $u$, starting where we left off in parameter space, in order to refine the search.  This approach is suggested by Fig.~\ref{costofR}, which shows the that the cost function becomes smoother with large coupling.  Provided the time interval is not too long, as discussed previously, $u$ can be reduced all the way to zero without significantly impacting the fit between model and data.  When this is not the case, there will be an (approximate) lower limit to $u$ below which synchronization is lost.  It is also possible to include $u$ in the cost function (Eq.~(\ref{costru})), and have it be reduced as part of the optimization process, although, as will be shown, this must be set up carefully in order to get the best result.  

The second approach we have utilized for this numerical optimization problem we call the `constrained method'.  In this method, one treats the values of the model state variables $\y(t_n)$ for all even $n$ on an equal footing with the parameters $\p$.  The differential equations are discretized to get equality constraints relating $\y(t_n)$ to $\y(t_{n+2})$, effectively forcing the optimization algorithm to find values which satisfy those differential equations.  The discretization process involves evaluation of the equations at the midpoint of the full time-step, i.e. the points $t_n$ where $n$ is odd (the details of this ``Hermite-Simpson" process can be found in Ref.~\cite{DRC}).  As before, we were able, for our particular data set, to make the choice that the half time-step for the model was equal to time-step for the data.  As a result, when we are using $N+1$ experimental data points in our analysis, this method needs to minimize the cost function in a space of dimension $D (1 + N/2) + L$, where $D$ is the number of variables and $L$ is the number of parameters.  When we allow the coupling $u$ to depend on $t_n$, this will add another $1 + N/2$ to this total.  Thus this dimension is typically about 2000 when we use $N+1=1001$ data points, i.e. the space we are optimizing over is \textbf{extremely} large.  We have successfully used up to $N+1=10001$ data points in the calculation.  We implemented the constrained method using the constrained nonlinear optimization software package SNOPT with the optimal control interface SNCTRL~\cite{SNOPT}. This is available publicly~\cite{snctrl}. We have used a symbolic mathematics code (Matlab or Mathematica) to provide the derivatives of all equality constraints and of the cost function.

The output of the numerical optimization using either method is an estimate of all the parameters along with an estimate of all of the unobserved state variables $y_2(t_n), y_3(t_n), ..., y_D(t_n)$ over the time interval of data presented to the model. For the Colpitts experiments, this gave us an estimate of the parameters not held fixed (see below) as well as $V_{CE}(t_n), I_L(t_n)\;;\;n = 0, 1, 2, ..., N$ over the interval from $t_0=0$ to $t_N = N \Delta t$. From the values of $V_E(t_N), V_{CE}(t_N), I_L(t_N)$ and the parameters we can use the differential equations to predict the temporal development of the oscillator.

As stated previously, we often wish to reduce the coupling to zero, at which point the rms error in fitting the data will be a good measure of our success in finding the correct parameters.  However, if it is impossible to reduce the coupling to zero, perhaps because of the excessive length of the data set, we can use the information from either method to ask an important consistency question about our estimation results by examining the magnitude of the ratio
\be 
R^2(t) = \frac{F_E(t)^2}{F_E(t)^2 + [u(t)(V_E(t) - V'_E(t))]^2},
\label{consistent}
\ee
where $F_E(t) = dV'_E /dt$ without the coupling term (see Eq.~(\ref{coupled})). If $R(t)$ is near unity over the time segment where we estimate parameters and state variables, the estimation is consistent as the coupling term is small compared the the dynamics in $F_E(t).$ If, however, $R(t)$ departs from unity, then the accuracy of any synchronization $\x(t) \approx \y(t)$ is due to the coupling. If $R(t) \approx 1$, we call the model consistent with the data. If not, especially if $R(t)$ is near zero, we call the model inconsistent with the data.

\section{State and Parameter Estimation using the Chaotic Colpitts Circuit as a test system}
As we have described, we operated the Colpitts circuit in a setting where chaotic trajectories for the states as a function of time were generated. We recorded $V_{CE}(t_n), V_{E}(t_n), I_L(t_n)$ at intervals of $\Delta t = 10$ $\mu$s for 100 ms ($t_n = n \Delta t, n=0, 1,..., 10^4$).  We then (unless stated otherwise) used the first 10 ms of the sampled $V_E(t_n)$ data as input into our estimation procedures.  As indicated earlier, we held fixed several parameters: 
$C_1, V_{cc}, V_{ee},$ and $R_{ee}$. There remained six parameters to determine from the 10 ms of $V_E(t_n)$ data: $C_2, L, R, I_s, V_T, \beta_F$.  Instead of directly treating the very small quantity $I_s$ as a parameter, we defined  $V_0 = -V_T \log \frac{I_s}{I_0}$, where $I_0 = 1 $ mA, and searched on $V_0$.   

We measured all values of the circuit elements directly, and we found the transistor parameters by measuring the I-V curves, $I_C(V_{BE})$ and $I_B(V_{BE})$, so that we could compare to the parameter estimates we obtain.  The resolution of the $V_E(t)$ measurement was $\Delta V =0.01$ ~V.  If the only source of uncertainty in the experiment was due to resolution of the measurement, we would expect an rms error (square root of the cost function, Eq.~(\ref{CostFn1})) of $\sqrt{C} = \Delta V/\sqrt{12} \approx 0.003$ ~V.  In both methods described below, the actual rms error between the model and the data comes out to be about four times larger, suggesting there are other sources of error.  

\subsection{With Time Independent Coupling }

\begin{figure}
\begin{center}
\includegraphics[width=1.0\columnwidth]{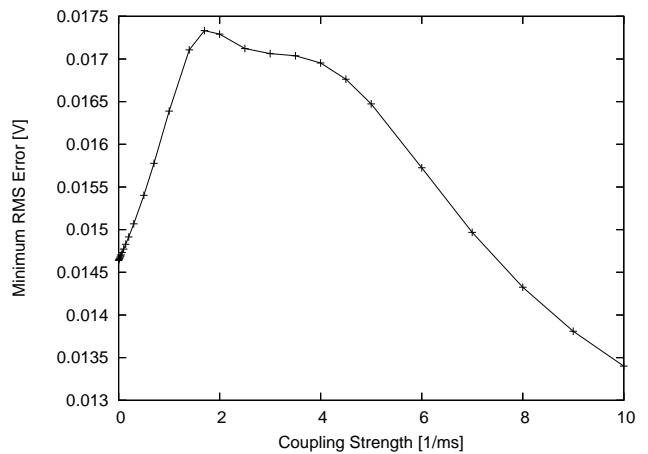}
\caption{Optimized rms error as a function of fixed, time independent coupling strength using 10 ms of real data and analyzed with the standard model.  The result is anomalous -- the initial increase in error with increasing coupling was unexpected.    Note that the coupling strength is not dependent on time-step for these results but rather is fixed at the values given on the horizontal axis.  Note also that the parameters are re-optimized for each point on the curves to reduce the rms error as much as is possible.}
\label{uError2}
\end{center}
\end{figure}

Here we use the DataSync software to implement the initial value method with time independent coupling $u$.  The response system is now a model of the circuit with six unknown parameters plus the three unknown initial conditions $V_E(0),V_{CE}(0),$ and $I_L(0)$. Exploring the behavior of the rms error in fitting the data as a function of coupling strength $u$ and find a {\bf very anomalous result} shown in Fig.~\ref{uError2}.  We expected the error to decrease with increasing coupling strength.  This is in fact the case for $u > 1.7$.  However, for $u$ between 0 and 1.7 the opposite is true -- the error rises and with a very significant positive slope even near $u=0$.

Recall that when we wish to have the optimization process reduce the value of the coupling $u$ towards zero, we would normally include a term containing the coupling strength in the cost function as in Eq.~(\ref{costru}).  However, if the coupling is restricted to the anomalous range between zero and the value where the rms error starts to decrease, then the coupling term {\bf may be omitted} from the cost function entirely.  We tested this using the initial value method with the coupling restricted to the range from 0.0 to 1.0 (1/ms), and with all parameters set initially to values that were either 1.4 or 0.7 times their best values and with the coupling started at 1.0.  The optimization was successful and the coupling ended up at about $3 \times 10^{-9}$.

If a wider range of coupling is desired then the value of $\eta$ that appears in the cost function must be chosen large enough to counteract the downward slope that occurs for large coupling.  If we let $\epsilon (u)$ represent the optimal rms error as a function of $u$, then the optimal cost function is given by $C(u) = \epsilon(u)^2 + \eta^2 u^2$.  Note that optimal means with parameters adjusted to minimize the cost for the particular choice of $u$.  If this function has a minimum for positive $u$ then the optimization process could get stuck there.  To look for a minimum we set the derivative equal to zero resulting in the equation:  $\eta^2 = (-d\epsilon/du) (\epsilon/u)$.  We can eliminate all such minima if we choose $\eta^2$ so that
\be 
\eta^2 > (-d\epsilon/du) (\epsilon/u),
\label{no_min}
\ee
for all u in the desired range.  In that case $u$ should tend to go to zero.

In the present case we can estimate $d\epsilon/du$ from Fig.~\ref{uError2}.  We found that for $u$ in the range from 0.0 to 10.0 the minimum value for $\eta^2$ should be about $2.6 \times 10^{-6}$.  Again we tested this with the initial value method.  With $\eta^2$ set to $2 \times 10^{-6}$ the optimization got stuck in a minimum near $u=6$, but when we increased $\eta^2$ to $3 \times 10^{-6}$ it no longer got stuck and successfully completed the optimization.  For this run $u$ ended up at about $1 \times 10^{-6}$.

Normally when starting to work on a new problem $d\epsilon/du$ will not be known in advance, so we would have to guess based on our knowledge of the problem and previous analyses of similar problems.  The difficulty in selecting $\eta$ can, of course, be avoided entirely by requiring $u$ to take on a series of values ending with zero.

\begin{figure}
\begin{center}
\includegraphics[width=1.0\columnwidth]{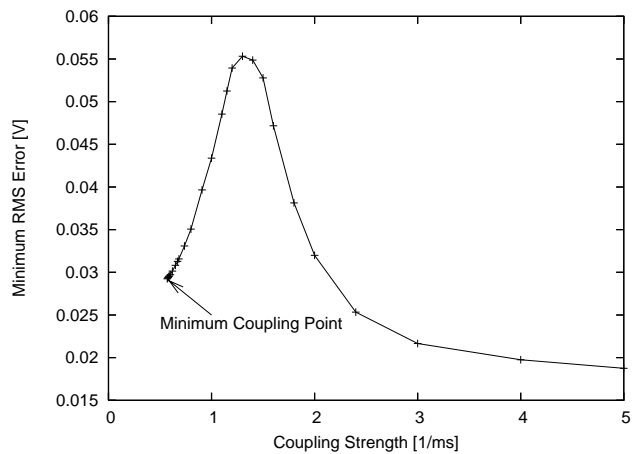}
\caption{Optimized rms error as a function of fixed, time independent coupling strength using 60 ms of real data and analyzed with the standard model.  Due to the length of the time series it is not possible to reduce the coupling strength to zero without a loss of synchronization.  The points were found by moving downwards slowly in $u$ and re-optimizing after each step.  It is not possible to continue the curve below the marked end point  -- instead the optimization is forced to jump to a very unfavorable state with at least 10 times the rms error.}
\label{uError6k}
\end{center}
\end{figure}

As mentioned previously, if we wish to be able to reduce the coupling to zero, we must ensure that the time interval $t_N$ over which we are analyzing the data is not too large.  In particular we must require that $t_N \lambda$ is not very large, where $\lambda$ is the largest positive Lyapunov exponent (0.35 for our data).  When $t_N$ is 10 ms as it is for most of our results there appears to be no problem in reducing the coupling to zero.  However if we increase $t_N$ to 60 ms this is no longer the case.  When we try to reduce the coupling in that case we reach a value below which it becomes impossible to maintain a good fit to the data.  This is illustrated in Fig.~\ref{uError6k}.  To create the curve in this figure, we start by optimizing at a fairly large value of coupling, e.g. $u=10$.  We then reduce $u$ in steps, each time starting the parameters with the values obtained in the previous step.  At each step we re-optimize the parameters and make sure that the rms error is only slightly different from the previous step.  If the step taken is too large the error settles on a value that is much larger than the previous step, often by a factor of 10 or more.  When we are near or below $u=1.0$ this process starts becoming exceedingly difficult.  We must take smaller and smaller steps and the optimization process becomes slower and slower.  Soon it becomes impractical to make any more progress.  For the case shown in Fig.~\ref{uError6k} this occurs around $u=0.57$, which is the lowest point shown.  {\bf Thus for this data-set we cannot reduce the coupling to zero.}  Our choices are: (a) use the results obtained at the minimal coupling $u=0.57$, (b) redo the analysis using a smaller time interval so that we can reduce the coupling to zero, (c) redo the analysis using a time dependent $u(t)$ as discussed in the next section, (d) break up the time interval into smaller subintervals, each of which is short enough to allow zero coupling.  The last case would be classified as a ``multiple shooting" method.  Each subinterval would have its own initial conditions to be determined, so there would be an increase in the total number of unknown parameters.  But this would be offset by the benefit of having more data points available

The parameters calculated with time independent coupling and using the shorter time series of 10 ms are compared to the measured values in Table \ref{tab:Params}.  The parameter estimates mostly come out near the measured values, with the exception of the transistor parameters $\beta_F$ and $V_0$.  We explore this discrepancy later with an improved transistor model.     

We also compute a value for the uncertainty of the parameters.  This was calculated by doing exactly the same fitting procedure, with the same initial guesses for parameter values, over nine different 10 ms segments of the same data set.  Since all the segments of data came from the same circuit, we expect all the parameter estimates to come out almost the same, but the initial condition estimates to be different.  The column labeled 'Uncertainty' in Table \ref{tab:Params} is the standard deviation of the mean of each of the parameters over the nine different 10 ms segments of data.  This is just a measure of the uncertainty from the numerical fitting process.  There is an additional uncertainty presumably due to experimental noise and other effects which are not included in the model.

\begin{table}
\begin{center}
\begin{tabular}{|c|r|r|r|r|c|}
\hline
Name & DataSync& Uncertainty & SNOPT & Measured & Units\\
\hline
$C_2$ & 6.98 & 0.03 & 7.02 & 7.23 & [$\mu$F] \\
\hline
$L$  & 12.28 & 0.02 &12.2& 11.74 & [mH] \\
\hline
$R$ & 40.38 & 0.01 & 40.0& 39.3 & [$\Omega$] \\
\hline
$V_0$ & 0.663 & 0.004& 0.661 & 0.63 & [V] \\
\hline
$V_T$  &  25 & 1& 25.0& 27 & [mV]\\
\hline
$\beta_F$ & 74 & 4 & 72.0 &180 & [1]\\
\hline
\end{tabular}
\caption{\label{tab:Params} Parameters found by the two methods and the measured values.}
\end{center}
\end{table}

Once the parameters are estimated with the fitting procedure, the model can be used to make predictions.  One simple prediction is the location of the fixed point.  The fixed point location is calculated using the estimated parameters (but substituting the measured value of $R$, since the resistance was increased to make the fixed point stable), and also can be directly measured from the circuit by making $R$ large enough to make the fixed point stable.  The measured and calculated values are shown in Table \ref{tab:FixedPoint}.  The values predicted using the fitted parameters are within 5\% of the measured values.  The third column in the table contains predictions from an improved model that will be explain later in the paper. 

\begin{table}
\begin{center}
\begin{tabular}{|c|r|r|r|r|}
\hline
Name & Measured& Prediction 1& Prediction 2 & Units\\
\hline
$V_E^{(0)}$ &  -0.696 & -0.723 & -0.702 & [V]\\
\hline
$V_{CE}^{(0)}$ &  2.24 & 2.31 & 2.25 & [V]\\
\hline
$I_L^{(0)}$ &   11.12 & 11.02 & 11.16 & [mA]\\
\hline
 
\end{tabular}
\caption{\label{tab:FixedPoint} Fixed point values.  The first column are the values measured directly from the circuit.  The second column is the calculation of the fixed point using the fitted parameters using the standard model.  The third column is the calculation of the fixed point using the fitted parameters from the improved model.}
\end{center}
\end{table}

\subsection{With Time Dependent Coupling}
Now we set the coupling $u$ to a function of time $u \to u(t_n)$, so the strength of the coupling (or control) to stabilize the synchronization manifold varies over the attractor. We use the SNOPT software as a way to implement the constrained method for solving the numerical optimization problem. As mentioned previously, for 10 ms of data we have space of approximately 2000 dimensions to search. As SNOPT is used regularly for optimization in spaces of high dimension, this was well within its tested regime of validity. We presented the software the data $V_E(t_n)$ at 1000 points, provided it with initial guesses and ranges for the six parameters, and set the initial guesses for all the state variables and coupling variables to zero.  

\begin{figure}
\begin{center}
\includegraphics[width=1.0\columnwidth]{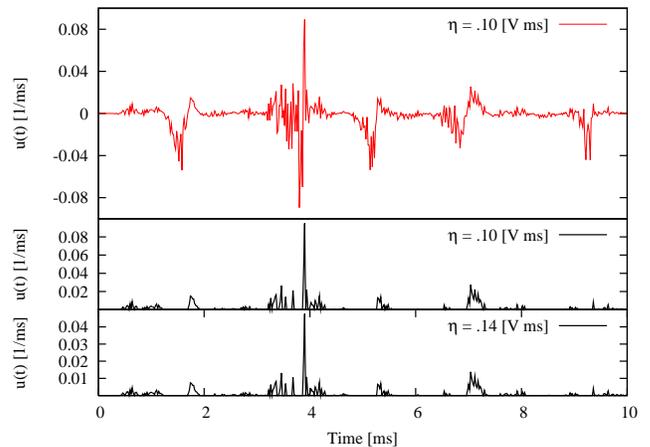}
\caption{Time dependent coupling $u(t)$ for $\eta = .10$ [V ms] (middle) and $\eta = .14$ [V ms] (bottom). The structure is very similar in both cases, but the magnitude scales approximately as $u(t) \propto \eta^{-2}$.  When $u(t)$ is allowed to be positive or negative, additional structure becomes visible(top). }
\label{Spikes}
\end{center}
\end{figure}

In the previous section we noted that when the coupling strength was below the peak in Fig.~\ref{uError2} there would be a tendency for it to move towards zero even with $\eta$ set to zero.  For time dependent coupling we find that there are subintervals in which the coupling will tend to decrease to the lower boundary (usually zero) and others in which it will tend to increase often going to the upper boundary (recall that we must require that $u \Delta t <1$ for numerical stability).  With a nonzero $\eta$ there emerges a set of peaks, shown in Fig.~\ref{Spikes}  whose amplitudes decrease with increasing $\eta$.  These peaks are not random, but are robust features for a given data-set.  The optimal rms error $\epsilon$ and the optimal cost $C$ are now (in a continuous time picture) functionals of the coupling strength $u(t)$.  Optimal here means the value obtained when the cost has been minimized over the parameter space for the particular choice of $u(t)$.  For a time interval from 0 to $\tau$, the cost can be expressed as:
\be 
C[u] = \epsilon[u]^2 + \frac{\eta^2}{\tau} \int_0^\tau u(t)^2 \,dt.
\label{cost}
\ee
We can solve for $u(t)$ by taking the functional derivative and setting it equal to zero:
\be 
\frac{\delta C[u]}{\delta u(t)} = 2 \epsilon [u] \frac{\delta \epsilon [u]}{\delta u(t)} + 2 u(t) \frac{\eta^2}{\tau} =0.
\label{deltaC}
\ee
We define $\epsilon_0$ to be the value of $\epsilon[u]$ for $u(t) = 0$ and we define $\kappa_0 (t)$ to be the functional derivative of $\epsilon[u]$ for $u(t) = 0$.  For sufficiently small $u(t)$ we may use these to solve for $u(t)$:
\be 
u(t) \approx - \frac{\tau \epsilon_0}{\eta^2} \, \kappa_0 (t).
\label{ufunc}
\ee
Thus we find that \textbf{the structure observed in $u(t)$ is proportional to the functional derivative of the rms error with respect to the coupling strength}.  The equation also shows that \textbf{the amplitude of the observed structure in $u(t)$ varies inversely with} $\eta ^2$ as can be seen in Fig.~\ref{Spikes}.  We usually set the lower limit on $u(t)$ to be zero, but when this is instead set to a negative value we can see additional structure corresponding to times when $\kappa_0 (t)$ is positive.  This is also shown in Fig.~\ref{Spikes}.

We found that $u(t)$ is strongly influenced by noise and this is responsible for the jagged appearance of $u(t)$.  To explore this we generated $u(t)$ using simulated data with added noise.  We then repeated this with the same data but using a different noise sequence of the same rms amplitude.  This produced a dramatic change in the structure demonstrating that the noise played a critical role.  We then repeated the test again, this time using the original noise sequence but with double the rms amplitude.  In that case we found that the structure remains essentially unchanged but the amplitude of that structure is increased by a factor of four, i.e. it is proportional to the square of the rms noise amplitude.

It is also interesting to see what happens if we change the form of the coupling term in the cost function.  In particular one can use a term that is linear such as $\xi u(t)$, where $\xi$ is a constant, resulting in a cost functional of the form:
\be 
C[u] = \epsilon[u]^2 + \frac{\xi}{\tau} \int_0^\tau |u(t)| \,dt.
\label{lincost}
\ee
If we take the functional derivative again we get
\be 
\frac{\delta C[u]}{\delta u(t)} = 2 \epsilon [u] \frac{\delta \epsilon [u]}{\delta u(t)} + \frac{\xi}{\tau}.
\label{lindeltaC}
\ee
Since this doesn't explicitly contain $u(t)$ we can't set it to zero and solve as before.  Note, however, that (in the small $u(t)$ limit) the right hand side will be positive when $\xi > - 2 \tau \epsilon_0 \kappa_0(t)$.  In these cases $u(t)$ will go to zero (the presumed lower boundary for $u(t)$).  For those values of $t$ for which the opposite is true, $u(t)$ will move upwards, either to the upper boundary, or to a value where our approximation breaks down because $u(t)$ is no longer small.  Further, one may expect that there will be a critical value of $\xi$ above which $u(t)$ will be zero for all $t$.

Note that in all of the cases above, we have been discussing the optimized value of $u(t)$.  In the process of getting to that optimal solution, $u(t)$ may temporarily take on much larger values as determined by the optimization algorithm.

In addition to noise, the structure in Fig.~\ref{Spikes} may be partially related to the varying relationship between the direction associated with the (scalar) data variable $V_E$ and the Lyapunov direction vectors (in Ref.~\cite{Brown} see discussion starting on page 2796), particularly direction of maximum stretching of the attractor that is associated with the largest Lyapunov exponent.  At certain locations on the attractor these directions are nearly parallel and so the coupling there is relatively effective in achieving synchronization.  This becomes increasingly important for long data-sets, particularly those which violate the condition of Eq.~(\ref{IntervalLimit}).  \textbf{These cases are also ones for which there may be a significant advantage for time dependent coupling.}  We previously examined a case for which the time interval was 60 ms, compared to 10 ms for most of our results.  In the results, shown in Fig.~\ref{uError6k}, we were unable to reduce the time independent coupling below 0.57.  At this point, the rms error was 0.0292, considerably higher that the value 0.0146 obtained in a shorter data-set.  Running this data-set with a time dependent $u(t)$ we were able to reduce this to 0.0161.  The average $u(t)$ is now much less than 0.57, but it contains a number of peaks with height of order 1, for which, at the time when they occur, the primary Lyapunov vector is relatively parallel to the coordinate axis associated with the measured experimental data variable.  When using the linear form of the coupling term in the cost function, a similar result is found except that the few remaining peaks are very tall and extremely narrow.  For $\xi = 0.001$ there are only 13 peaks in $u(t)$ with an average height of about 6 and an average width of only slightly more than one full time-step.  Everywhere else it is zero.  One might reasonably assume from this, that for many problems with a good choice of $\xi$ there will be little impact on the dynamics other than to maintain synchronization.  Of course, if $\xi$ is too large the synchronization will be adversely effected while if it is too small the number of peaks in $u(t)$ will become excessive and detrimental as it attempts to correct for every noise bump in the data.

\begin{figure}
\begin{center}
\includegraphics[width=0.80\columnwidth]{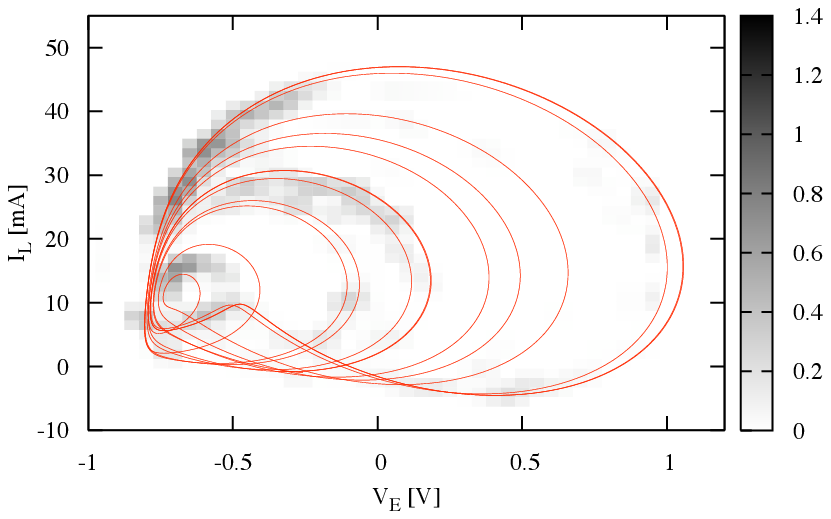}
\includegraphics[width=0.80\columnwidth]{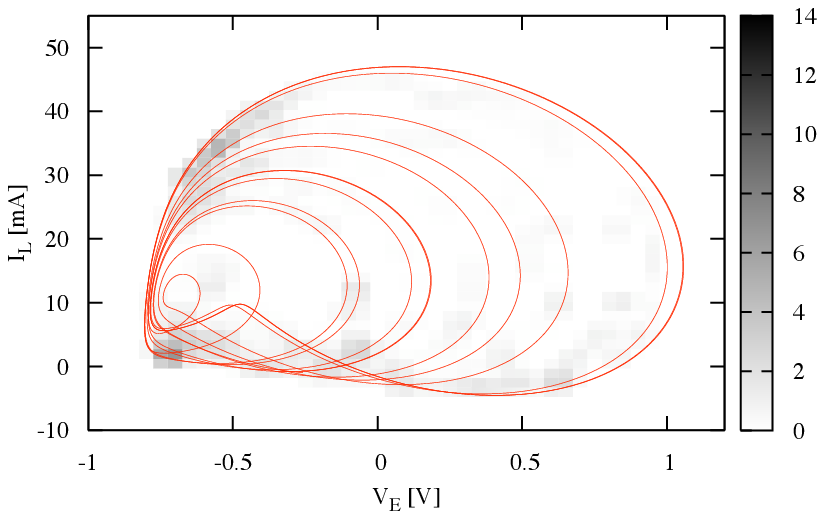}
\caption{Localization of peaks of $u(t)$ in phase space.  Phase space was divided into 40 x 40 x 1 cells ( 40 in the $I_L$ and $V_E$ directions, 1 in the $V_{CE}$ direction), and the average value of $u(t)$ in each cell is indicated by the shading. The calculated phase space orbit is overlaid for reference. The coupling was either to $V_E$ ( {\bf Top Panel}), or to $V_{CE}$ ({\bf Bottom Panel}).  In both cases $\eta = .01$ [V ms] and 1800 data points where used.}
\label{Pixels}
\end{center}
\end{figure}

In Fig.~\ref{Pixels} we show that the peaks of the residual $u(t)$ are localized in phase space.  (Here we are using 1800 points in the data set and a quadratic term in the cost function). We divided the phase space into cells and shade in each according to the average value of $u(t)$ within each cell.  We do the same thing a second time, but using $V_{CE}(t)$ as the data source instead of $V_E(t)$.  The areas where $u(t)$ is largest are now in different spots in phase space.

In the cost function Eq.~(\ref{costru}) we explored various values of the weight $\eta$ and report in Table~\ref{tab:Params} the values of the parameters for $\eta = 0.1$ $[V][ms]$.  If $\eta$ is chosen too small, $u$ will become large and the coupling term will end up dominating the dynamics in Eq.~(\ref{coupled}).  This means that $V'_E(t)$ will be forced to follow $V_E(t)$ even if the parameter values or the model itself is wrong.  On the other hand, if $\eta$ is chosen too large, $u$ will be driven toward zero and synchronization may not occur.  In either extreme limit of $\eta$ very small or $\eta$ very large the ratio $R(t) \approx 1$, because either $(V_E(t) - V'_E(t))$ or $u$ will become very small in Eq.~(\ref{consistent}).  Therefore care must be taken to chose a value of $\eta$ that allows the two effects to be balanced. 

\begin{figure}
\begin{center}
\includegraphics[width=1.0\columnwidth]{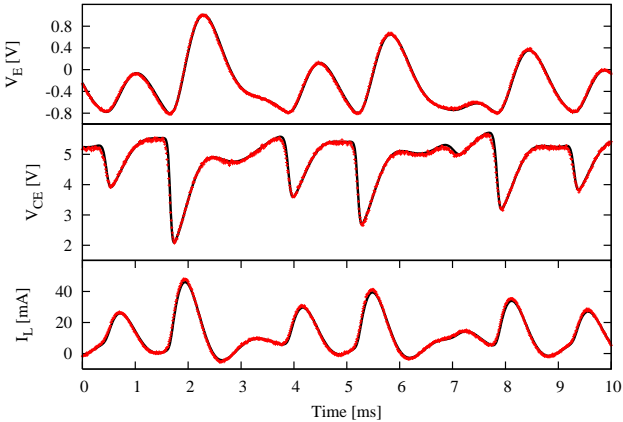}
\caption{Using 1000 measurements  of $V_E(t)$ over 10 ms (dots in top panel), we estimate the other two state variables, $V_{CE}(t)$ and $I_L(t)$ (solid lines). Also shown for comparison is the measurement of $V_{CE}(t)$ and $I_L(t)$ (dots in middle and bottom panels).}
\label{stateEst}
\end{center}
\end{figure}

The parameter search methods provided consistency for estimates of the six parameters as displayed in Table \ref{tab:Params}. In 
Fig.~\ref{stateEst} we display both the observed values of the state variables $[I_L(t), V_E(t), V_{CE}(t)]$ and their estimated values from the use of the constrained method. Only $V_E(t)$ as observed was presented to the estimation algorithm, yet the estimation of the two unobserved state variables $I_L(t)$ and $V_{CE}(t)$ is also quite accurate. 

\begin{figure}
\begin{center}
\includegraphics[width=1.0\columnwidth]{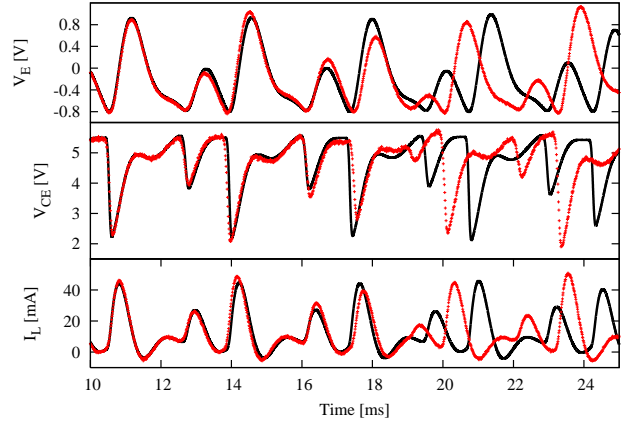}
\caption{Using the first 10 ms of data for $V_E(t)$, sampled every 10 $\mu$s, we estimated the parameters and the other state variables $V_{CE}(t)$ and $I_L(t)$ over the interval 0 $\leq t \leq$ 10 ms. Then using the equations of motion with estimated parameters and the values of $V_E(t = 10 $ms) (as measured) and $V_{CE}(t = 10 $ms) and $I_L(t = 10$ ms) (as estimated), we predicted the three state variables for 10 ms $< t <$ 25 ms.  The predictions (solid lines) are compared to the measured data (dots).  Initially the prediction matches the measurement, but the two diverge after about 8 ms because the dynamics are chaotic.}
\label{statePred}
\end{center}
\end{figure}

From the output of the estimation program we have an estimate of the values of the three state variables at $t = 1000 \Delta t = 10\,ms$, and using these as an initial condition for the model of the Colpitts circuit along with the values of the estimated parameters (Table \ref{tab:Params}), we predicted the behavior of the circuit for $t > t_{1000}$. These predictions are displayed in Fig.~\ref{statePred}. Each shows that the predictions from the estimated state of the oscillator are accurate for about 8 ms or so beyond the point in time $t = 10$ ms where the state estimate was made. This is the kind of prediction one finds when a dynamical system is chaotic as small errors in parameters or initial conditions are exponentially magnified in the behavior of the orbit because of the positive Lyapunov exponents of the system. What is important here is that we have been able, through our estimation methods, to accurately enough evaluate the state variables that were not observed that our model predictions of the experimental system show high accuracy.

\begin{figure}
\begin{center}
\includegraphics[width=1.0\columnwidth]{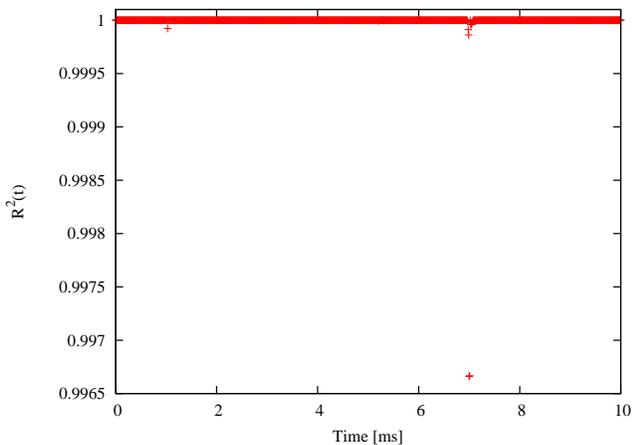}
\caption{The consistency check $R^2(t)$ defined in Eq.~(\ref{consistent}). Since $R^2(t) \approx 1$ the model is consistent with the data. In this calculation $\eta = .1$ [V ms].
}
\label{Roft}
\end{center}
\end{figure}

Because we use a coupling to the synchronization manifold as part of our estimation procedure we need to ask whether the synchronization we observe is the result of large coupling.  At the end of the optimization process we calculate the ratio $R^2(t)$ from Eq.~(\ref{consistent}), and see that $R^2(t) > .99$ at all times as shown in Fig.~\ref{Roft}.  This means that the coupling has a negligible impact on the dynamics, as it should, and that our model is consistent with the data.

\subsection{Effects of Model Deficiencies and Results for the Gummel-Poon Model}
The Ebers-Moll model is only an approximate model of a real transistor.  This causes interesting challenges for the parameter calculation process and makes the Colpitts oscillator more interesting as a test problem.  In a general problem, it may often be the case that the value of the parameters themselves will be of interest.  But in the process of trying to optimize the output of an inexact model sometimes the deficiencies in the model may be partially compensated for with parameter values that are dramatically shifted away from their true values.  In the current problem this may have occurred with the current gain parameter $\beta_F$, for which we obtained a value of about 72 by minimizing the cost function compared to a measured value of about 180.  Another anomaly was observed in regards to the capacitor $C_1$. For the results shown in Table \ref{tab:Params}, $C_1$ was fixed to its known value while other parameters were calculated.  In principle, $C_1$ could also be obtained by optimization, but when we attempted this it was found that for this particular problem, shifting $C_1$ away from its true value has only a very weak influence on the optimal cost value. Furthermore, some deficiency in the model causes the optimization to prefer a drastically incorrect value for $C_1$ while maintaining a correct value for the series equivalent capacitance, $C_{eq} = 1/(1/C_1 + 1/C_2)$.  It seems that $C_{eq}$ has matters more than $C_1$ or $C_2$ separately, which is reasonable because the fundamental frequency is given by $f_0 = 1/(2 \pi \sqrt{L C_{eq}})$. 

To see if we could reduce or eliminate these effects, we tried adding some additional parameters to our model equations using a more accurate model of the transistor known as the Gummel-Poon model \cite{GP}.  The full model involves an additional 10 parameters (excluding capacitive effects which we assume are negligible in our application).  These include modeling internal resistance between the emitter of the transistor on the chip and the package lead which we connect to the rest of the circuit.  There is a similar resistance between the collector and its package lead.  The base resistance is assumed to vary with base current according to a particular formula including three adjustable parameters.  There is also an assumed rolloff to the current gain involving one parameter.  The ÒEarly voltageÓ parameter compensates for changes in base thickness with collector voltage.

It turns out that a significant improvement in the cost function can be achieved by merely adding the emitter resistance into the model equations.  Because of the presence of this resistor, $-V_E$ is no longer the true base-emitter voltage, but instead it has a slightly higher voltage $-\bar{V_E}$.  The equations for the transistor currents (Eqs.~(\ref{IC}) and (\ref{IB})) are now functions of $\bar{V_E}$. The modified voltage $\bar{V_E}$ is a function of $V_E$ which must be found numerically by finding the solution to:
\be
\bar{V_E} = R_E I_s (1 + \frac{1}{\beta_F}) \exp \left({\frac{-\bar{V_E}}{V_T}}\right) + V_E.
\ee
Using only $R_E$ in addition to our previous parameters, the rms error in matching the data was reduced from about 0.0146 to about 0.0102 in volts.  In addition, the value of $\beta_F$ increased from 72 to 179, i.e. remarkably close to the measured value of 180. We also tried other parameters from the Gummel-Poon model in various combinations including the full set, but found no significant additional improvement in the cost function.  Although the value of $\beta_F$ was improved, the problem with $C_1$ remains, so perhaps the Gummel-Poon model still retains significant deficiencies.  These results were obtained using our first method (initial value), and are summarized in Table \ref{tab:GP}.  The parameters where also used to make a prediction of the fixed point, shown in Table \ref{tab:FixedPoint}.

\begin{table}
\begin{center}
\begin{tabular}{|c|r|r|c|}
\hline
Name & Fit & Measured & Units\\
\hline
$C_2$ & 7.08 &  7.23 & [$\mu$F] \\
\hline
$L$  & 12.00 &  11.74 & [mH] \\
\hline
$R$ & 39.71 &  39.3 & [$\Omega$] \\
\hline
$V_0$ & 0.637  & 0.63 & [V] \\
\hline
$V_T$  &  26 & 27 & [mV]\\
\hline
$\beta_F$ & 179 &180 & [1]\\
\hline
$R_E$  &  .23 &  &[$\Omega$]\\
\hline
\end{tabular}
\caption{\label{tab:GP} Parameters found with the improved transistor model, which includes emitter resistance $R_E$, compared to the measured values}
\end{center}
\end{table}

\subsection{Comparison of Models}

\begin{figure}
\begin{center}
\includegraphics[width=1.0\columnwidth]{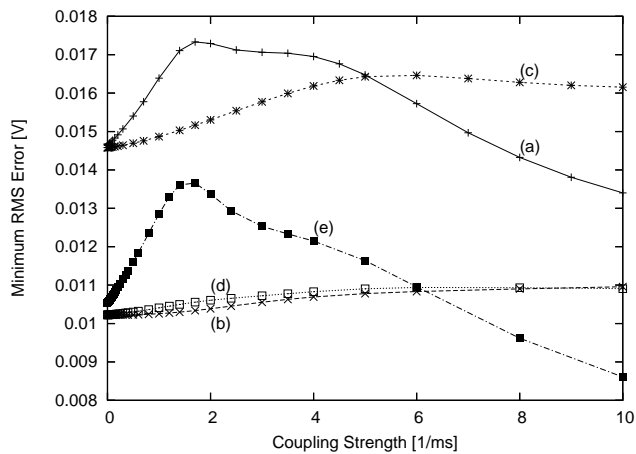}
\caption{Optimized rms error as a function of fixed, time independent coupling strength.  (a) Real data analyzed with standard model. (b) Real data analyzed with improved model. (c) Simulated data generated from standard model plus noise, and analyzed with standard model. (d) Simulated data generated from improved model plus noise, and analyzed with improved model. (e) Simulated data generated from improved model, and analyzed with standard model. }
\label{uError}
\end{center}
\end{figure}

In order to better understand the effect of coupling on the optimization process, we again looked at how the coupling strength effects the value of the rms synchronization error in fitting the data. These are shown in Fig.~\ref{uError}.  Curve (a) (taken from Fig.~\ref{uError2} and included here for reference) shows the case where we are using experimental data and use our standard model (Eqs.~(1-5)), while (b) shows this data represented using our improved model that include the effects of an internal emitter resistance $R_E$.  Note that in both cases the seemingly logical assumption that increasing the coupling strength will always decrease the rms error is violated.  Note that curve (b) starts at a significantly lower error than (a), a strong indication that it is a better model.  In curves (c) and (d) we are looking to see whether the anomalous behavior of (a) and (b) are due to inaccuracies in the model or whether the presence of additive noise can produce a similar effect.  Here we used simulated data with the same parameter values found in the experiment to which Gaussian noise was added.  In case (c) the simulation used Eqs.~(1-5), i.e. it did not involve the use of an internal emitter resistor, while in (d) that resistor was included.  In both cases the noise level was adjusted to produce about the same rms error for $u=0$ as is found in the curves (a) and (b).  Note that presence of noise can produce a positive slope effect, but seems to do so in a much less dramatic fashion than curve (a).  Unlike curves (a) and (c), curves (b) and (d) seem much more similar to each other.  This is perhaps an indication that the model used in those cases is a much better one.  In the final curve (e), the simulated data was generated using $R_E$, but then it was analyzed with the model that lacks this resistor.  No noise was added in that case, so the rms error is entirely due to a lack of ability of the model to fit the simulated data.  Note the similarity to curve (a), both peak at $u$ approximately 1.7.  It starts at a lower rms error than (a) because the real data does include some added noise.

\begin{figure}
\begin{center}
\includegraphics[width=1.0\columnwidth]{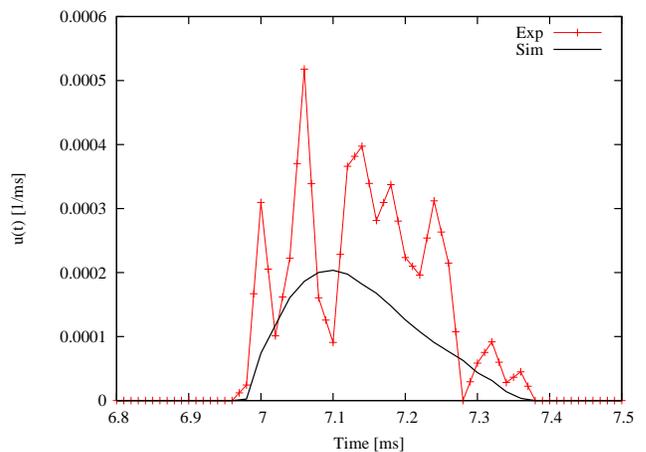}
\caption{Optimized coupling strength $u(t)$ as a function of time with $\eta = 1.0$.  (a) Jagged curve from experimental data, analyzed without emitter resistance.  (b) Smooth curve for simulated data with parameters matching the experimental data that was generated using an emitter resistance but analyzed without (imperfect model). }
\label{Sim_Re_U_of_t}
\end{center}
\end{figure}

Model imperfections are yet another cause of the structure seen in the time dependent coupling function $u(t)$.  We can demonstrate this by reanalyzing the simulated data of case (e) above, using our constrained method with time dependent coupling.  The result shows much smoother features in $u(t)$ than are generated by noise in the data.  In Fig.~\ref{Sim_Re_U_of_t} we show just a portion of the $u(t)$ function showing an overlay of the result for the simulated data with that of the experimental data.  Note that the experimental data appears to show the same feature but with noise superimposed.  \textbf{This result suggests another way to identify imperfection in a model is to look for these larger scale smooth features in $u(t)$.}

\section{Discussion}

We have developed a method, based on synchronization, for finding unknown parameters and predicting the future state of an actual chaotic experimental system, from limited measurements.  This problem is difficult for several reasons: the dynamics are chaotic, there are many parameters and state variables to search over, we have only very limited measurements of the experimental system, and most importantly, we do not have an exact description of the experiment in our models. 

Because of the chaotic nature of the dynamics, the cost function is an especially rough surface in the space of parameters and initial state values of the response system.  This makes it difficult to locate the global minimum, which represents the point where the model system parameters match the data system parameters.  This problem can be made dramatically easier by adding a term to the dynamics of the model system which couples the response system to the driver system.  For sufficient coupling strength, this often has the effect of changing the largest conditional Lyapunov exponent from positive to negative, and stabilizes the synchronization manifold.  This effect can be observed by noticing that the coupling smoothes out the cost function (Fig.~\ref{costofR}).

Even with the coupling, it is still difficult to search for a global minimum in a space with many dimensions.  To perform the search we used a few different well-established optimization methods \cite{SNOPT, nr, Brent}.  This is mainly just a practical problem of efficiency.  We did this with time independent coupling and time dependent coupling with almost identical results.  Not surprisingly, the process takes much longer with time dependent coupling, because there are many more degrees of freedom to be optimized.

We explored two different types of coupling.  The simpler type is time independent coupling, which works quite well on our test problem.  If the time series being analyzed is short enough then coupling can be reduced all the way to zero at the end of the process.  Otherwise it is reduced to the minimum level needed to maintain synchronization.  We found an interesting effect where the optimal rms error in fitting the model to the data initially increases as the coupling strength is increased.  This was very unexpected since the coupling is \textbf{pulling} the model variable \textbf{towards} the data.  As the coupling is further increased, the error reaches a peak and then begins to decrease, so that for large coupling the behavior is what would be expected.  The result is likely an artifact of the noise in the data, i.e. the coupling term is sometimes responding more to this noise than to the actual synchronization error.

The other type of coupling is with time dependent $u(t)$ which is optimized like the other parameters to yield the lowest cost function.  This makes the problem more complicated, but gives us some additional, possibly useful, information such as the peaks in the residual $u(t)$ which are robust and are localized in phase space.  This structure was also shown to be highly dependent on how $u(t)$ is included in the cost function -- a term that is quadratic in $u$ leads to a very rich structure that mirrors the sensitivity of the synchronization error to the coupling as a function of time; a term that is linear in $u$ leads to a few very sharp narrow peaks in the residual $u(t)$, with the vast majority of time-steps having a $u$ that is exactly zero.  We suggest that this form of coupling may be best for maintaining synchronization for long data sets while causing minimal effect on the dynamics.

By doing tests with simulated data we saw that there are three separate causes that contribute to the structure of the residual coupling $u(t)$.  
\begin{itemize}
  \item One is due to inaccuracies in the model.  In regions where the model is inaccurate, the coupling increases to pull the model toward the data.  This information could be useful for finding out where the model needs improvement.  
  \item The second is due to noise.  We saw that $u(t)$ changes significantly if only the particular sequence of random numbers used to generate the noise is changed, but there are persistent features.  
  \item The third cause is related to the direction of the primary Lyapunov vector relative to the direction associated with the coupling variable.  This becomes relevant, even with a perfect model, when the time series is long enough to require coupling to counteract positive Lyapunov exponents.
\end{itemize}

We also described a new method of coupling that is particularly useful for sparse data, i.e. data that has a time-step that is much too large to be used as the time-step for the model because it would lead to significant numerical error.  This coupling method is applied discretely at the exact locations in time where the data points reside.

We considered the important effect of imperfect models.  In the Colpitts case, most of the circuit components behave in a well known nearly ideal way, and can be very accurately described by linear differential equations.  The exception is the transistor.  Even though it is moderately well described by the Ebers-Moll model, the slight inaccuracy that is present is sufficient to cause significant errors in the estimated values of some parameters.  This can be verified by including an additional term in the equations -- one that is part of the more detailed Gummel-Poon model \cite{emoll, GP}.  In many other cases that one might choose to study, the models are not as well established or as accurate and there will likely be similar problems but more severe.  Our main interest in this simple case is that it provides a way of testing methods, that then can be applied to the harder cases. 

In this case we were then able to make improvements to the transistor model by adding the emitter resistance $R_E$, but if we had no information besides $V_E(t)$ we might have declared the original model good enough and stopped there.  With that issue in mind, we then explored how the rms error of the fit depends on coupling strength (Fig.~\ref{uError}). 

We proposed a test that can be done with just a measurement of one of the dynamical variables ($V_E(t_n)$ in this case), and one prospective model.  The main idea is that if the experiment is accurately described by the prospective model, then data simulated using the prospective model should be indistinguishable from the real experimental data.  Here we are not comparing the real data directly to the simulated data, but instead comparing the results of the fitting process (performed using either the same model or a different model) applied to both real data and simulated data.  We showed in Fig.~\ref{uError} that when simulated data using the improved model (with noise added) is fed into the fitting algorithm, the output (curve d) looks almost the same as the output of feeding the real data into the same fitting algorithm (curve b).  Although there are many details of this test we do not understand yet, it seems to be useful way of checking the quality of a prospective model.

Improvements to a model, when available, provide another test of our analysis -- they are important when they cause a significant change in one or more of the estimated parameters or in the values of the variables that are inaccessible in the experiment. 

In a perspective on the ideas in this paper, we note that we addressed three problems:
\begin{itemize}
  \item when comparing data from an experimental source and a model where the time series has chaotic oscillations, one must regularize the search mechanism (minimizing a cost function) to avoid the irregularities in parameter and initial condition space so the impediments to the search of multiple local minima are removed,
  \item one must select numerical methods that give the parameters and state variables over the observation period,
  \item and one must develop good models to compare to the data, and be aware that seemingly small defects in a model can have a very strong impact on the results obtained.
\end{itemize}
The methods we have described, including the two possible choices for the second item that we explored, impact the quality of the model in an interesting fashion: the techniques can be used to explore the quality of a model, as we demonstrated in the investigation of two somewhat different models for the nonlinear transistor element in the Colpitts circuit.

\section*{Acknowledgments}
This work was partially funded by the U.S. Office of Naval Research MURI grant (ONR N00014-07-1-0741).  We would like to thank Erik Lindberg of the Technical University of Denmark for useful discussions about the Colpitts oscillator.  We would also like to thank Philip Gill and Elizabeth Wong for helpful discussions about SNOPT and SNCTRL software.


\end{document}